\documentstyle[aps,twocolumn,prl,psfig]{revtex}
\begin{document}
\draft

\title{Orbital Ordering Induced Antiferromagnetic Spin Fluctuations\\
 in the Ferromagnetic State of 50 \% Hole-doped Manganites:\\
Pr$_{1/2}$Sr$_{1/2}$MnO$_{3}$ and Nd$_{1/2}$Sr$_{1/2}$MnO$_{3}$}
\author{H. Kawano,$^{1,2}$ R. Kajimoto,$^{3}$ H. Yoshizawa,$^{3}$\\
Y. Tomioka,$^{4}$ H. Kuwahara,$^{4}$ \cite{kuwadd} and Y. Tokura$^{4,5}$}
\address{$^{1}$Institute of Physical and Chemical Research (RIKEN), Wako, Saitama 351-0198, Japan\\
$^{2}$Solid State Division, Oak Ridge National Laboratory, Oak Ridge, Tennessee 37831\\
$^{3}$Neutron Scattering Laboratory, ISSP, University of Tokyo, Tokai, Ibaraki 319-1106, Japan\\
$^{4}$Joint Research Center for Atom Technology (JRCAT),
Tsukuba, Ibaraki 305-8562, Japan\\
$^{5}$Department of Applied Physics, University of Tokyo, Bunkyo-ku, Tokyo 113-8656, Japan }

\date {\today}

\twocolumn[\hsize\textwidth\columnwidth\hsize\csname@twocolumnfalse\endcsname

\maketitle

\begin{abstract}
We report inelastic neutron scattering results on the spin dynamics in the 50 \% hole-doped manganites, Pr$_{1/2}$Sr$_{1/2}$MnO$_{3}$ and Nd$_{1/2}$Sr$_{1/2}$MnO$_{3}$. In the paramagnetic phase, these systems exhibit a ridge-type diffuse scattering, indicating a two-dimensional ferromagnetic (FM) correlation. With decreasing temperature, the systems  enter the FM state, but simultaneously they develop A-type antiferromagnetic spin fluctuations.  The spin wave dispersion in the FM state is strongly anisotropic. These behaviors can be consistently interpreted from a viewpoint of the underlying orbital ordering.
\end{abstract}
]

%\section{Introduction}

Recent experimental and theoretical studies on Colossal Magnetoresistance (CMR) manganites have demonstrated that the one-electron bandwidth ($W$) and orbital ordering play an important role in determining their physical properties such as spin and lattice structures, transport properties, charge ordering, spin dynamics, {\it etc.}\cite{cho98,mae98,miz97,kos97,kaw97}. For example, we have studied three perovskite manganites, Pr$_{1-x}$Sr$_{x}$MnO$_{3}$ and Nd$_{1-x}$Sr$_{x}$MnO$_{3}$ with $x \sim 1/2$, and demonstrated that, instead of the well-known CE-type spin and charge order, some of manganites exhibit a metallic layered A-type antiferromagnetic (AFM) state\cite{kaw97}, in which ferromagnetic (FM) layers stack antiferromagnetically [See  Fig. 1].  The change of the magnetic structure from a CE-type to an A-type can be interpreted as an effect of the widening of $W$.  Based on the detailed study of the lattice and spin structures, we pointed out that these systems may have the $d_{x^{2}-y^{2}}$- type orbital order. This orbital order favors the FM spin alignment within orbital ordered layers, but favors the AFM stacking between them, leading to the A-type AFM order. Furthermore,  we  proposed that this orbital order may introduce a two-dimensional character in both the magnetic and transport properties in the A-type AFM state of the nearly 50 \% hole-doped manganites.  We have recently demonstrated that the spin wave (SW) stiffness constants in the A-type AFM state of Nd$_{0.45}$Sr$_{0.55}$MnO$_{3}$, indeed, exhibit a clear anisotropy between the two directions within and perpendicular to the orbital ordered FM layers\cite{yos98}.  Furthermore, very recently, it was reported that the resistivity of this compound also shows a clear directional anisotropy \cite{kuwxx}. Because of the metallic conductivity in the A-type AFM phase, one can conclude that the charge ordering is not formed in these systems and that the anisotropy in the magnetic as well as transport properties observed in this phase should be solely attributed to the ordering of $d_{x^{2}-y^{2}}$ orbitals.

It is interesting to point out that, judging from the crystal structure of the Pr and Nd compounds, such a unique $d_{x^{2}-y^{2}}$-type orbital order may persist even in their  FM state, and it could influence their magnetic as well as transport properties.  If this is the case, it may give a unique opportunity to investigate the influence of the orbital ordering on their physical properties separately from the charge ordering.  The choice of the present two compounds provides further advantage to identify the effects of the orbital ordering.  Because the orbital order is formed along the different directions [See Fig. 1], it should exhibit a different directional dependence.  With this difference in mind, we have performed inelastic neutron scattering measurements in the paramagnetic (PM)  and  FM metallic states of 50 \% hole-doped Pr and Nd manganites.

In this paper we report that, reflecting the underlying $d_{x^{2}-y^{2}}$-type orbital order, the spin fluctuations in these systems show several  anomalous behaviors.  In particular, the spin correlations in these systems are two-dimensional in the PM phase, and they turn to possess {\it A-type AFM spin fluctuations} even in the FM state.  In addition, they show clear directional dependence.

%\section{experimental procedure}

%\subsection{neutron scattering experiments}

Neutron scattering measurements were performed with triple axis spectrometers GPTAS-4G and HER-C11 installed at the thermal and cold-guide tubes at the JRR-3M research reactor, JAERI, Tokai, Japan. The GPTAS spectrometer was operated with $k_{f}=2.57$\AA$^{-1}$ and 40'-80'-40'-80' collimators, while the HER with $k_{i}=1.55$\AA$^{-1}$ and open-80'-80' collimators from monochromator to detector, respectively.  The latter yields an energy resolution of $\sim 0.23$ meV (FWHM).  The measurements were performed with the same crystals used in our previous studies.  The details on the sample preparation and the sample quality have been already reported\cite{kaw97,tom95,kuw95}.  The single crystal samples were aligned with the $(h, l, h)$ scattering plane. The SW profiles were measured around the respective A-type AFM Bragg points, $Q=(1/2, 0, 1/2)$ or $(1, \pm 1, 1)$, while the FM spin fluctuations were measured around the nuclear Bragg reflection $Q=(1,0,1)$, respectively.

%\section{experimental results}

\begin{figure}[htb]
\centering \leavevmode
\psfig{file=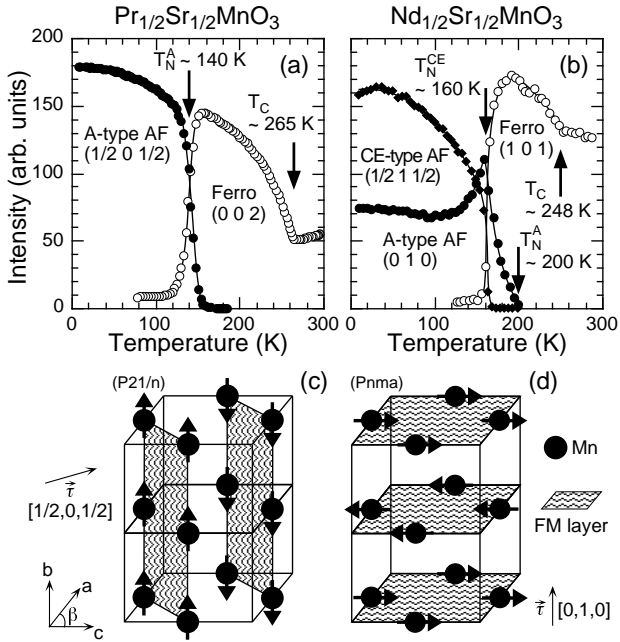,width=0.95\hsize}
\vspace{2mm}
\caption{Upper panels: Temperature dependence of the FM and AFM Bragg reflections in Pr$_{1/2}$Sr$_{1/2}$MnO$_{3}$ and in Nd$_{1/2}$Sr$_{1/2}$MnO$_{3}$.  Lower panels: A-type magnetic structures observed in two systems, and their propagation vectors $\vec{\tau}$. The $d_{x^{2}-y^{2}}$-type orbitals lie within the hatched planes.}
\label{fig1_magstr}
\end{figure}

The magnetic as well as transport phase diagrams for Pr$_{1-x}$Sr$_{x}$MnO$_{3}$ and Nd$_{1-x}$Sr$_{x}$MnO$_{3}$ systems have been reported by Tomioka et al.\cite{tom97} and by Kuwahara et al.\cite{kuw97}.  For $x=1/2$, these two systems show first order magnetic phase transitions either from the FM to A-type or to CE-type state for Pr and Nd compounds,  respectively. The temperature dependence of the magnetic Bragg intensity in both systems is summarized in the upper panels of Fig. 1.  In the present study, however, we have newly discovered that a very weak A-type AFM reflection appears below $T_{N}^{A} \sim$ 200K in the Nd compound, and that this is a second order transition.

The crystal structures of these compounds belong to either the orthorhombic Pnma or monoclinic P21/n symmetry. (Detailed lattice parameters were given in Ref. \onlinecite{kaw97,kaj98}.)  The important feature of these crystal structures is that the MnO${_6}$ octahedra in both systems consist of two long and one short O-Mn-O bonds, and that two long O-Mn-O bonds are always contained in the shaded layers illustrated in Fig. 1.  Consequently, the $d_{x^{2}-y^{2}}$-type orbitals lie within these planes, and the propagation vector $\vec{\tau}$ for the concomitant A-type AFM order is $\vec{\tau} =$ [101] and $\vec{\tau} =$ [010] for the Pr and Nd compounds, respectively.  Hereafter, we employ a terminology of "{\it layer}" and "{\it plane}" to indicate this orbital-ordered plane shown in Fig. 1, although these compounds do not have a conventional layered structure.

%\subsection{constant $E$ profiles}

\begin{figure}
\centering \leavevmode
\psfig{file=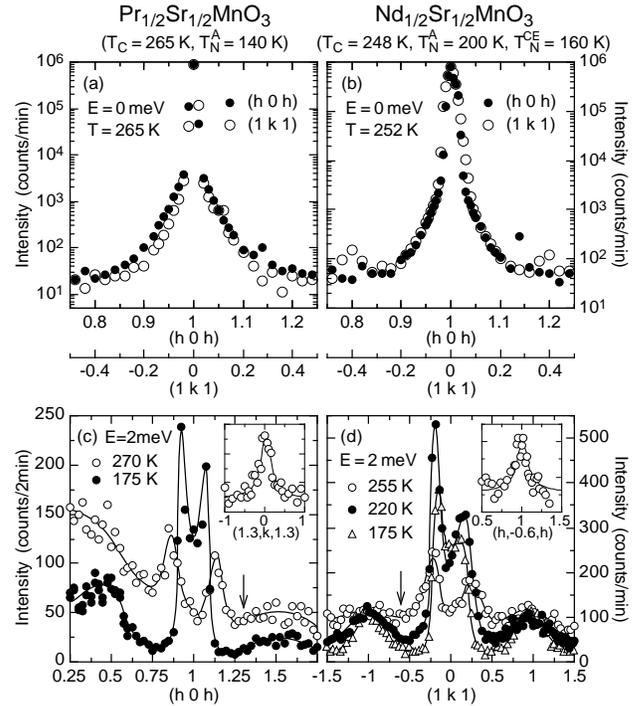,width=0.95\hsize}
\vspace{2mm}
\caption{Constant-E scan profiles observed at $E=0$ (upper panels) and 2 meV (lower panels).  (Insets) Profiles observed along the perpendicular to the 2d rod-like scattering at 270 K and 255 K for the Pr and Nd compounds, respectively. The arrows in the bottom panels indicate the positions where the profiles were measured.}
\label{fig2_constE}
\end{figure}

In Fig. 2 we summarize the wave vector dependence of PM spin fluctuations measured at $E= 0$ and 2 meV, respectively.  These data were taken at the HER spectrometer with an energy resolution of $\sim 0.23$ meV (FWHM).  The profiles of the $E= 0$ meV component in the upper panels consist of a strong nuclear Bragg component at the center and additional wing scattering.  The diffuse component at the wing is practically identical for the $h$ and $k$ directions, indicating that the $E= 0$ meV diffuse component is isotropic.  In sharp contrast to this,  the $E= 2$ meV component depicted in the lower panels is clearly anisotropic.  For example, if we examine the data of the Nd compound in Fig. 2(d), the profile observed in the PM state at 255 K is almost flat along the $k$ direction except for two spurious peaks at $k \sim \pm 0.25$ \cite{com1}, while the profile along the $h$ direction is narrow. This indicates that the dynamical spin correlations are two-dimensional; the spins are strongly correlated ferromagnetically within the  planes, but are not correlated between the planes, reflecting the anisotropic feature of the $d_{x^{2}-y^{2}}$-type orbital order.  These results manifest that the orbital ordering has a crucial influence on the dynamical spin correlations in manganites even in the PM state. When decreasing $T$, a further intriguing feature develops at around the A-type AFM zone center, $Q = (1, \pm 1,1)$.  Namely, the profile observed at 220 K indicates that spins between the planes now exhibit AFM correlations even though the system is still in the FM phase.  Similar AFM spin correlations were also observed at 175 K below $T_{N}^{A}$.

For the Pr compound, we observed exactly the same behavior with the Nd compound except for the change of the direction of the spin correlations as shown in Fig. 2(c).  In this compound, the two-dimensional FM spin correlations develop within the $(101)$ planes, (consistent with the fact that the orbital order is formed within these planes), and A-type AFM correlations develop at around $Q = (1/2, 0, 1/2)$ and $(3/2, 0, 3/2)$.  The increase of intensity towards small $h$ in profiles is due to small-angle scattering.

%\subsection{SW dispersion relations}

\begin{figure}
\centering \leavevmode
\psfig{file=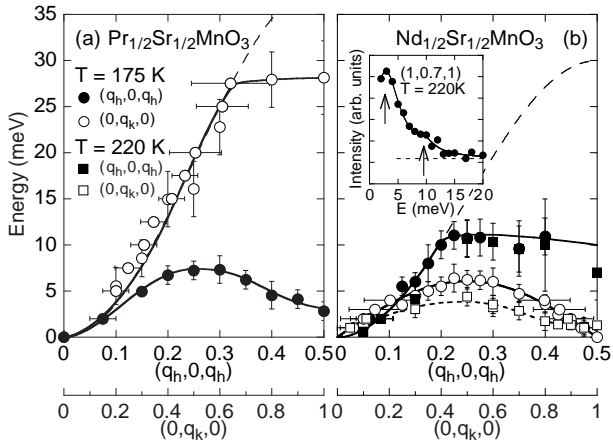,width=0.95\hsize}
\vspace{2mm}
\caption{Dispersion curves within and between plane directions. An inset shows a constant-$Q$ scan profile observed at $(1, 0.7, 1)$.  Solid curves are drawn as guides to the eye.}
\label{fig3_disp}
\end{figure}

In order to characterize the spin fluctuations in the FM phase, we determined the SW dispersions.  As shown in Fig. 3, there are several remarkable features in the SW dispersion relations within and between the planes.  First, it is evident that SW stiffness is different for the two directions.  For example, in the Pr compound, the SW energy along the [1 0 1] direction (perpendicular to the orbital-ordered plane) is lower than those along the [0 1 0] direction (an in-plane direction within the orbital-ordered FM planes) [See Fig. 1(c)].  We already reported that the same orbital order causes such anisotropy in the SW dispersion in the A-type AFM state\cite{yos98}.  Thus, an implication of the present results is that the orbital ordering causes same anisotropic SW dispersion even in the FM state, which should be otherwise isotropic.  

The second feature is the anomalous $q$-dependence of the dispersion curve along the in-plane direction.  In the Pr case, the SW dispersion curve deviates in the middle of the Brillouin zone from a single cosine-band type formula given by conventional linear SW theory which is indicated by the dashed curve. This curve is calculated for the simple FM Heisenberg model with nearest-neighbor couplings, and is identical to the one predicted from the double exchange model in the limit of large Hund's coupling by Furukawa\cite{noby96}.  This relation was reported to explain  well the observed SW dispersion relations in a relatively wide one-electron bandwidth ($W$) systems La$_{0.7}$Pb$_{0.3}$MnO$_{3}$\cite{per96} and La$_{0.7}$Sr$_{0.3}$MnO$_{3}$\cite{mar96}.  However, the deviation from the linear SW theory is serious for the Pr compound, and is even more striking for the Nd compound.  In the Nd case, the dispersion along the in-plane direction (along the [101] direction for the Nd compound) becomes practically flat beyond $q_{h} = 0.2$.  The inset of the Fig. 3(b) is the profile observed at (1, 0.7, 1).  There are two peaks at 3 and 9.5 meV due to the two magnetic domains.  One might suspect that a crystal electronic field excitation from Nd ions gives rise to the peak at 9.5 meV.  However, we can easily exclude this possibility because the profile measured at the zone center (1,0,1) exhibits no peak at 9.5 meV.  At present, the origin of the deviation is an open question.  For the present two compounds, they have almost identical $T_{C}$'s but different $W$, indicating that the difference of $W$ could explain to what extent the SW dispersion deviates from the linear SW theory.  A similar deviation was recently reported for the lower hole-doping compounds Pr$_{0.63}$Sr$_{0.37}$MnO$_{3}$, Nd$_{0.7}$Sr$_{0.3}$MnO$_{3}$, and La$_{0.7}$Ca$_{0.3}$MnO$_{3}$\cite{hwa98,Dai98}.

Thirdly, the dispersion relations perpendicular to the orbital-ordered layers also exhibit anomalous $q$-dependence.  In the Pr compound, the zone boundary (ZB) along the $(q_{h}, 0, q_{h})$ direction  in the FM phase is located at $q_{h} =  0.5$.  Surprisingly, however, it seems that the ZB is located at $q_{h} =  0.25$ and that the periodicity of the dispersion is doubled along this direction, although the SW energy at the ZB is still finite.  For the Nd compound, the dispersion along the $(0, q_{k}, 0)$ direction exhibits the same behavior, but the SW energy at the ZB with $q_{k} =1.0$ vanishes below $T_{N}^{A} \sim 200$ K.

To see detailed behavior of the ZB SW excitations, we show in Fig. 4 the $q$-dependence of the SW profiles near the ZB at two $T$'s above and below $T_{N}^{A} \sim 200$ K for the Nd compound.  At 220 K in the FM phase (right panel), the ZB profile shows a clear gap, but at 179 K below $T_{N}^{A}$ (left panel) the $q$-dependence indicates that the gap vanishes at the ZB of $q_{k} =1.0$. This behavior is consistent with the emergence of the A-type AFM Bragg reflections at $T_{N}^{A}$ [See Fig. 1(b)], and sheds light on the nature of the ordering process of two compounds.  Due to the orbital ordering, the dynamical spin fluctuations are anisotropic in the PM phase.  With decreasing $T$, the A-type AFM spin correlations are developed between {\it orbital-ordered planes} in the FM state, and they are progressively enhanced.  When further decreasing $T$, they yield a weak parasitic A-type AFM order in the Nd compound, but the narrow W favors the CE-type charge and spin order further, and it is finally formed below $T_{N}^{CE}$.  For the Pr compound which has a slightly wider $W$, on the other hand, the orbital-order induced A-type AFM spin fluctuations lead to a metallic A-type AFM order through a first order transition.

\begin{figure}
\centering \leavevmode
\psfig{file=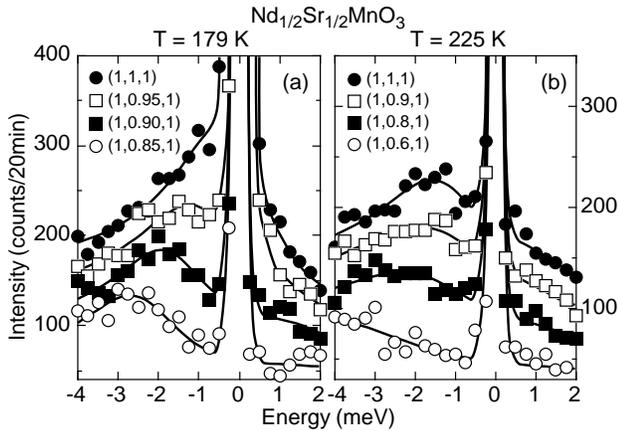,width=0.95\hsize}
\vspace{2mm}
\caption{The $q$-dependence of the constant $Q$ scan profiles observed above and below A-type AFM transition temperature $T_{N}^{A} \sim 200$ K.}
\label{fig4_T-dep}
\end{figure}

%\section{discussion}

Finally, we would like to discuss an issue of a FM central component.  Recently, it was pointed out that the CMR manganites which show large CMR phenomena have a strong central component at around their $T_{C}$ \cite{lyn96,fer98}.  The central component can be also observed in the PM state of the AFM manganites \cite{yos98,kaj98b,bao97}. Such a central component yields a spin diffusion constant with one order smaller energy scale to its spin stiffness constant\cite{yos98,kaj98b}.  The present data [Fig. 2] demonstrate that the correlation length of the central component is order of $\sim$100 \AA.  In addition to the central component, however, the system has an additional component which has a larger energy scale and shorter correlation length than those of the central component.  This component must correspond to conventional thermally-activated spin fluctuations, indicating the coexistence of slowly fluctuating large clusters and conventional spin fluctuations in the system.  The former slow fluctuation is very likely related to the CMR phenomena, while the latter dynamical component clearly manifests an influence of the underlying orbital ordering.

%\section{conclusions}

In conclusion,  we have found a number of interesting but unusual features in the spin dynamics of the 50 \% hole-doped manganites Pr$_{1/2}$Sr$_{1/2}$MnO$_{3}$ and Nd$_{1/2}$Sr$_{1/2}$MnO$_{3}$ in and above the FM phase. (1) The spin correlations in the PM state are two-dimensional.  (2)  The SW dispersion in the FM state is anisotropic.  (3) The SW dispersion along the in-plane direction shows a  deviation from a conventional linear SW theory.  (4)  AFM spin correlations are developed in the FM state.  We attribute the results (1), (2), and (4) to the underlying $d_{x^{2}-y^{2}}$-type orbital order, which yields the A-type AFM order in the low temperature phase. The directional dependence of the spin correlations observed in two systems is consistent with our interpretation that these spin fluctuations are driven by the specific orbital order.  These results demonstrate that the orbital ordering has a crucial influence on the spin correlations in the 50 \% hole-doped manganites, even in the PM and FM states.  The reason of the result (3) remains an open question, and further experimental and theoretical studies are needed.

% Acknowledgement
 
We thank Dr. J. A. Fernandez-Baca for valuable discussions and for a critical reading of the manuscript. This work was supported by a Grant-In-Aid for Scientific Research from the Ministry of Education, Science, Sports and Culture, Japan and by the New Energy and Industrial Technology Development Organization (NEDO) of Japan. The work at ORNL was supported by U.S. DOE under contract No. DE-AC05-96OR22464 with Lockheed Martin Energy Research Corp.

\end{document}